\documentclass[preprint,prb,amssymb,superscriptaddress,showpacs]{revtex4}
\usepackage{graphicx}
\usepackage{amsmath}

\begin{document}

\title{Nonuniversality of the interference quantum correction to
conductivity beyond the diffusion regime}
\author{A.\,V.\,Germanenko}
\affiliation{Institute of Physics and Applied Mathematics, Ural
State University, 620083 Ekaterinburg, Russia}

\author{G.\,M.\,Minkov}
\affiliation{Institute of Metal Physics RAS, 620219 Ekaterinburg,
Russia}

\author{A.\,A.\,Sherstobitov}
\affiliation{Institute of Metal Physics RAS, 620219 Ekaterinburg,
Russia}

\author{O.\,E.\,Rut}\dag
\affiliation{Institute of Physics and Applied Mathematics, Ural
State University, 620083 Ekaterinburg, Russia}

\date{\today}

\begin{abstract}
Results of numerical simulation of the weak localization in
two-dimensional systems in wide range of magnetic filed  are
presented. Three cases are analyzed: (i) the isotropic scattering
and randomly distributed scatterers; (ii) the anisotropic scattering
and randomly distributed scatterers; (iii) the isotropic scattering
and the correlated distribution of the scatterers.  It is shown that
the behavior of the conductivity beyond the diffusion regime
strongly depends on the scattering anisotropy and correlation in the
scatterer distribution.
\end{abstract}


\maketitle

Interference (or weak localization) quantum correction to the
conductivity arises from interference of electron waves scattered
along closed trajectories in opposite directions. An external
magnetic field applied perpendicular to the two-dimensional (2D)
layer  destroys the interference and suppresses the quantum
correction. It results in anomalous negative magnetoresistance,
which is experimentally observed in many disordered 2D
systems.\cite{AA85} Theoretically, this problem was studied only for
the case of random distribution of scattering centers and isotropic
scattering (comparative analysis of different calculation approaches
has been done in Ref.~\onlinecite{McPh}). As a rule, these
conditions are not fulfilled in real semiconductor structures. First
of all, the scattering by ionized impurities dominates often in
semiconductors at low temperatures. This scattering is strongly
anisotropic, in particular, in heterostructure with remote doping
layers. Besides, the impurity distribution is correlated to some
extend due to Coulomb repulsion of the impurity ions at the
temperatures of crystal growth.

In the present work the role of scattering anisotropy in the quantum
correction to the conductivity is investigated through a computer
simulation. It is shown that  the interference quantum correction is
nonuniversal beyond the diffusion regime, when it is determined by a
short closed paths with relatively small number of collisions. Its
behavior with temperature and magnetic field change depends on the
scattering details strongly.

\begin{figure}
\includegraphics[width=0.5\linewidth,clip=true]{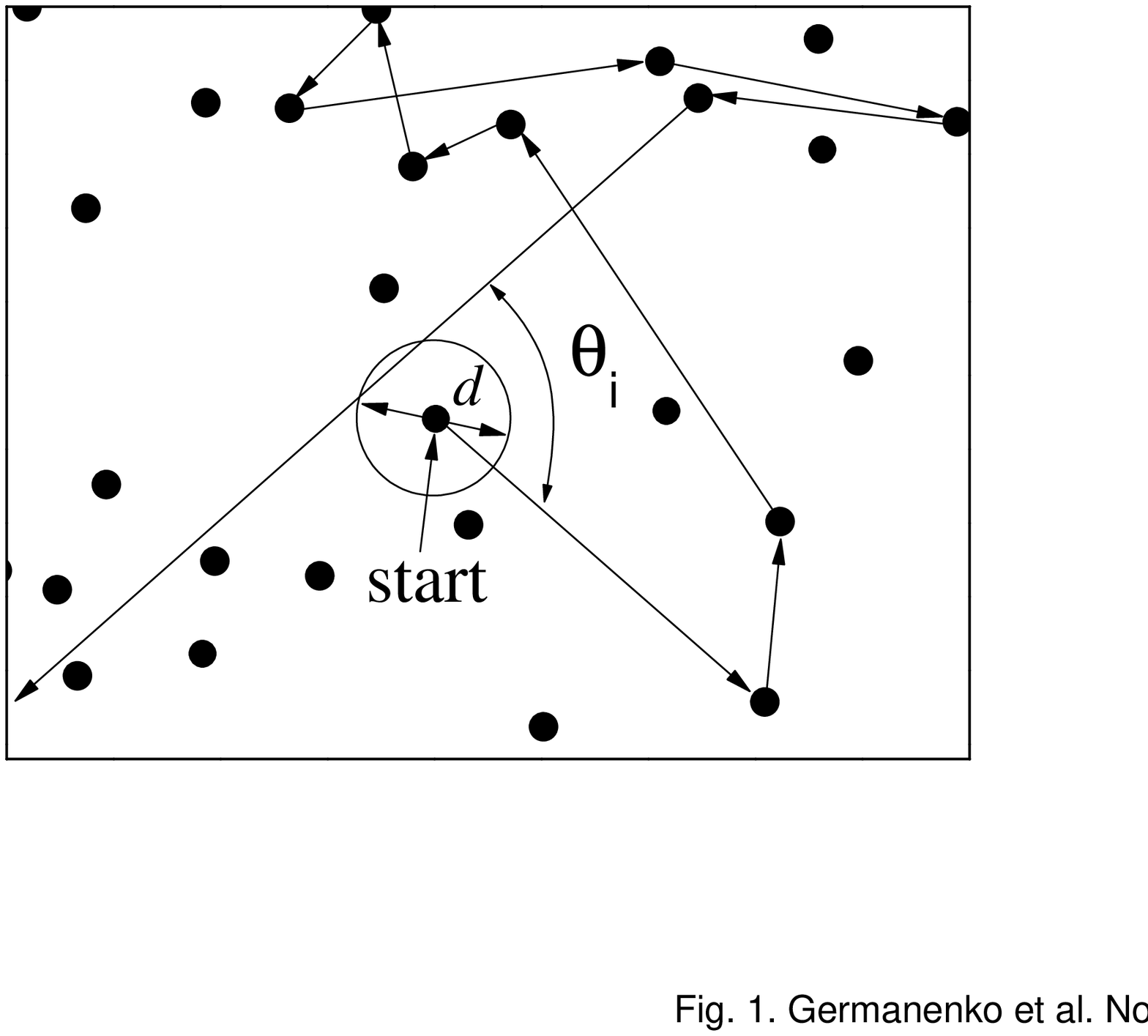}
\caption{Sketch of  model 2D system and fragment of the $i$-th
closed path near starting point. }\label{fig00}
\end{figure}

The weak-localization phenomenon can be described in the framework
of quasiclassical approximation which is justified under the
condition $k_F l\gg 1$, where $k_F$ is the Fermi wave vector, $l$ is
the mean free path. In this case the conductivity correction is
expressed through the classical {\it quasi}probability density $W$
for an electron to return to the area of the order $\lambda_F l$
around the start point \cite{gork,chak,dyak,gorn}
\begin{equation}
\label{eq1} \delta\sigma=-\sigma_0 \frac{\lambda_F l}{\pi} W = -
2\pi l^2 G_0 W,
\end{equation}
where $G_0=e^2/(2\pi^2\hbar)$, $\lambda_F=2\pi/k_F$,  and
$\sigma_0=\pi G_0 k_F l$ is the Drude conductivity. Prefix {\it
quasi-} means that $W$ takes into account the dephasing of
interfering waves caused by external magnetic field and inelastic
processes. Expression (\ref{eq1}) allows ones to calculate the
conductivity correction at any magnetic field and
temperature.\cite{schm}  To find  $W$ and $l$, we simulate the
motion of a particle over the 2D plane with scattering centers in
it. This technique was described in details in our paper,
Ref.~\onlinecite{our1}. Here is outline only. The plane is
represented as a lattice. The scatterers with a given cross-section
are placed in a part of lattice sites with the use of a random
number generator. A particle is launched from some random point,
then it moves with a constant velocity along straight lines, which
happen to be terminated by collisions with the scatterers. After
collision it changes the motion direction.  If the particle passes
near the starting point at the distance less than $d/2$ (where $d$
is a prescribed value, which is small enough), the path is perceived
as being closed. Its length and enclosed algebraic area are
calculated and kept in memory. The particle walks over the plane
until it escapes the lattice. As this happens one believes that the
particle has left to infinity and will not return. A new start point
is chosen and all is repeated. A fragment of one of the closed paths
near the starting point is shown in Fig.~\ref{fig00}.

\begin{figure}[t]
\includegraphics[width=0.7\linewidth,clip=true]{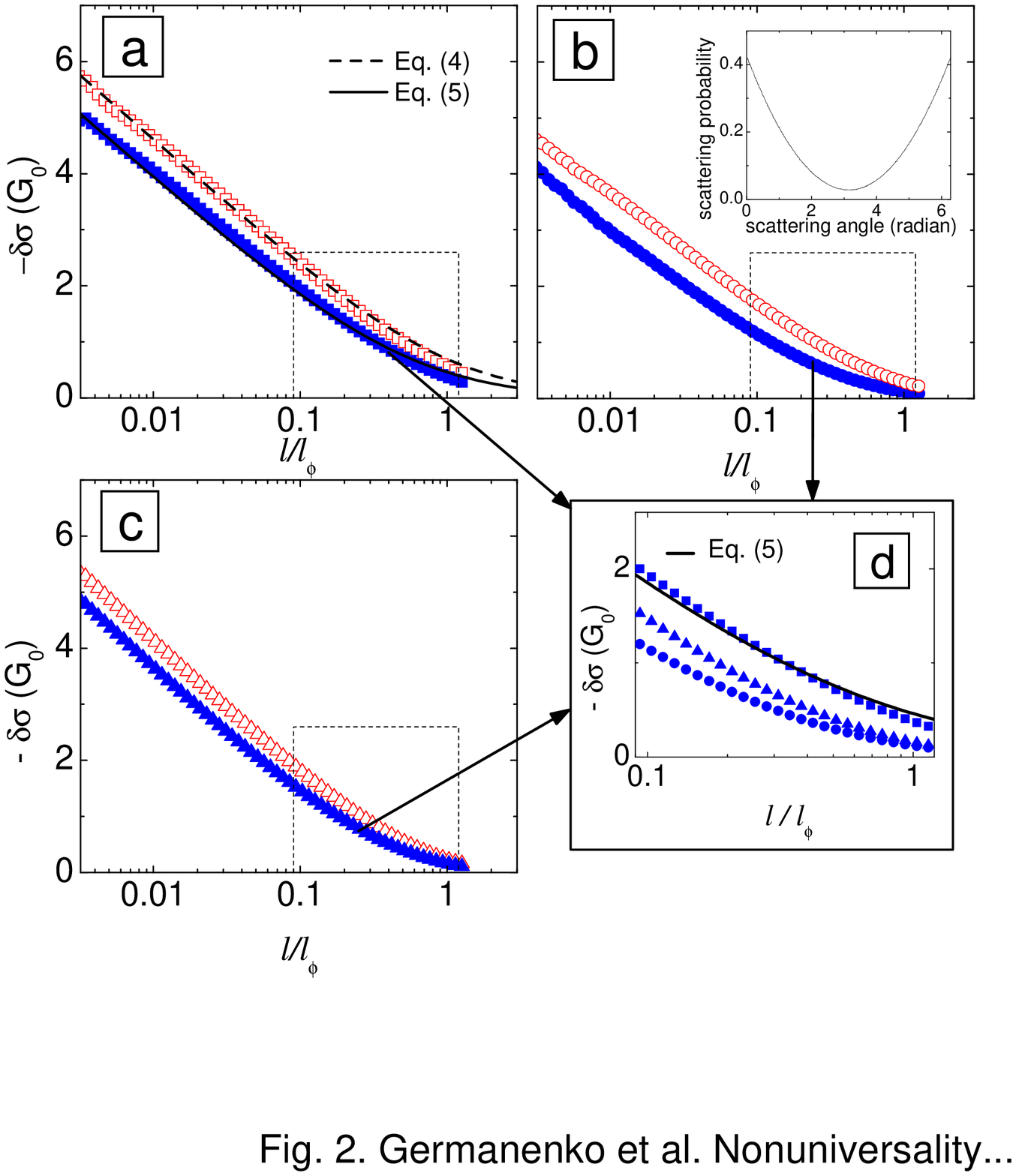}
\caption{The interference quantum correction  in zero magnetic field
as a function of $l/l_\phi$. (a) --  Scattering is isotropic, the
distribution of scatterers is random (system A). Dashed and solid
lines are the analytical results,  Eq.~(\ref{eq4}) and
Eq.~(\ref{eq41}), respectively. (b)~-- Scattering is anisotropic
with the angle dependence of the scattering probability shown in the
inset, the distribution of scatterers is random (system B). (c) --
Scattering is isotropic, the distribution of scatterers is
correlated (system C).  (d)~-- The same data are presented in
enlarged scale. Open and solid  symbols correspond to
Eqs.~(\ref{eq2}) and (\ref{eq3}), respectively.}\label{fig10}
\end{figure}

The simulation has been carried out for three systems. In  system A
the scatterers are distributed randomly, the scattering is isotropic
that physically corresponds to a short-range scattering potential.
In the system B the scatterers are distributed randomly and
scattering is anisotropic. The angle dependence of scattering
probability for this case is presented in the inset in
Fig.~\ref{fig10}(b). It is close to that for the heterostructure
with $\delta$ doped barrier in which impurities are spaced from the
2D gas. The anisotropy shown can correspond, for instance, to a
semiconductor heterostructure with impurity density of about
$10^{12}$~cm$^{-2}$ and a spacer thickness of $50$~\AA. Finally, the
system C is characterized by isotropic scattering, but the
distribution of scatterers is quite correlated. Namely, the nearest
neighbor distance is not less than 20 whereas the mean distance
between scatterers was about 30 (hereafter length and area are
measured in units of lattice parameter and its square,
respectively).

The parameters used in simulation are the following: $d=5$; the
number of launches $10^5..10^6$; the dimensions of the lattice is
$10^4\times 10^4$. The cross-section of the scatterers is equal to
7. The density of scatterers is such that the transport mean-free
path is about $120$ for all three systems.  For illustration, let us
set the lattice parameter equal to $5$ \AA. Then our model provides
an example of 2D system in which the transport mean-free path is
$l=600$ \AA\ and $B_{tr}\simeq 0.09$~T.

As shown in Ref.~\onlinecite{our1}, the value of $\delta\sigma$ for
the model systems can be calculated as follows
\begin{equation}
\label{eq2} \frac{\delta\sigma(b)}{G_0}=-\frac{2\pi l}{d\cdot
N}\sum_{i} \cos\left(\frac{bS_i}{l^2}\right) \exp \left(-
\frac{l_i}{l_\phi} \right),
\end{equation}
where summation runs over all closed trajectories among a total
number of trajectories $N$,  $b=B/B_{tr}$ is the magnetic field
measured in units of transport magnetic fields
$B_{tr}=\hbar/(2el^2)$, $S_i$ and $l_i$ stand for the algebraic area
and length of the $i$-th trajectory, respectively, $l_\phi$ is the
phase breaking length connected in reality with the phase relaxation
time $\tau_\phi$ through the Fermi velocity $v_F$, $l_\phi=v_F
\tau_\phi$ (not to be confused with the diffusion phase breaking
length $L_\phi=\sqrt{D\tau_\phi}$, where $D$ is the diffusion
coefficient).  Equation~(\ref{eq2}) takes into account only the
coherent backscattering correction to the conductivity. In order to
take into account the nonbackscattering processes, which are
important in the ballistic regime,\cite{gorn} each term in
Eq.~(\ref{eq2}) should be multiplied by the factor
$[1-\cos(\theta_i)]$,\cite{our1} where $\theta_i$ is the angle
between the first and last segments of the $i$-th closed path (see
Fig.~\ref{fig00}):
\begin{equation}\label{eq3}
\frac{\delta\sigma(b)}{G_0}=-\frac{2\pi l}{d\cdot N}\sum_{i} \left[
... \right] \left[ 1-\cos(\theta_i)\right].
\end{equation}
In what follows the results obtained from both Eq.~(\ref{eq2}) and
Eq.~(\ref{eq3}) will be considered.

In Fig.~\ref{fig10}, the results of our simulation for
$\delta\sigma(b=0)$ are plotted against  the ratio $l/l_\phi$ (or
that is just the same against $\tau/\tau_\phi$). Open symbols
correspond to the case when only  the backscattering processes are
taken into account, i.e., $\delta\sigma$ is calculated from
Eq.~(\ref{eq2}). Data presented by solid symbols are obtained from
Eq.~(\ref{eq3}) and, thus, include both back- and nonbackscattering
contributions. In the assumption of $\tau_\phi \propto T^{-p}$,
where $p>0$, this figure illustrates the temperature dependence of
the interference quantum correction to the conductivity.  The upper
solid line in Fig.~\ref{fig10}(a) shows the known theoretical
results for the backscattering contributions in weak
localization\cite{chak,gorn}
\begin{equation}\label{eq4}
  \frac{\delta\sigma}{G_0}=-\ln\left(1+\frac{\tau_\phi}{\tau}\right).
\end{equation}
The lower line is the results of Ref.~\onlinecite{gorn} where both
the backscattering and nonbackscattering contributions are taken
into account
\begin{eqnarray}
\frac{\delta\sigma}{G_0}&=& - \ln{\left(1+\frac{\tau_\phi}
{\tau}\right)} \nonumber \\
&+&\frac{1}{1+2\tau_\phi/\tau}\ln{\left(1+\frac{\tau_\phi}
{\tau}\right)}+ \frac{\ln{2}}{1+\tau/2\tau_\phi}, \label{eq41}
\end{eqnarray}

As is clearly seen the simulation and the theoretical data are
almost the same. Thus, one can believe that the simulation procedure
does not fail and for $l/l_\phi>3\times 10^{-3}$ our model system is
equivalent to an unbounded 2D system and the simulation gives
correct results.

Inspection of Fig.~\ref{fig10} shows that  the temperature
dependence of $\delta\sigma$ in the systems with the anisotropic
scattering (system B) and with the correlation (system C) is very
close to that for the system with isotropic scattering and random
distribution of scatterers (system A). In all three cases the
temperature dependence of $\delta\sigma$ is close to the logarithmic
one for low $l/l_\phi$ values, i.e., for low temperatures in real
systems. The only difference is the absolute value of the
interference correction for a fixed $l/l_\phi$ value. The relative
difference in the temperature behavior for different systems becomes
more pronounced at $l/l_\phi\gtrsim 0.1$ [see Fig.~\ref{fig10}(d)]
when the quantum correction is mainly determined by the closed paths
with a small number of collisions (so called ballistic regime).
Thus, already analysis of the data carried out in the absence of
magnetic field shows that the interaction correction feels the
details of scattering.

\begin{figure}
\begin{center}
\includegraphics[width=0.8\linewidth,clip=true]{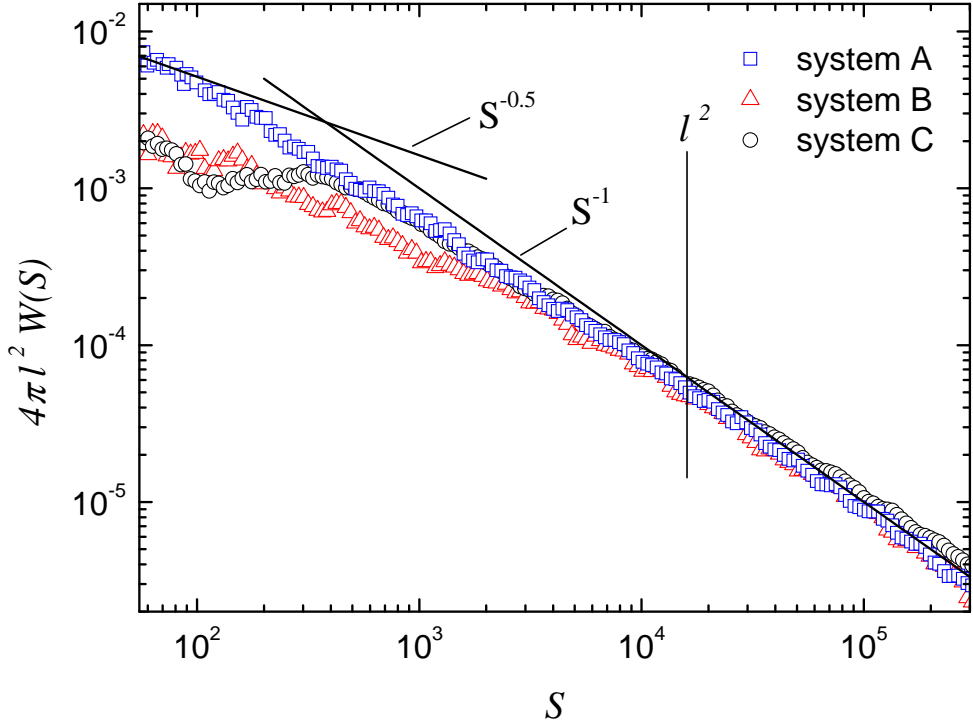}
\end{center}
\caption{The area distribution function of closed paths, obtained
from the simulation procedure. Solid lines are the asymptotes for
the purely ballistic and diffusive motion. Since $W(S)$ is identical
for $S>0$ and $S<0$, only the range of positive algebraic area is
shown.} \label{fig20}
\end{figure}

Before to discuss the magnetoconductivity let us  consider the area
distribution function of closed paths $W(S)$ because namely it
determines the magnetic filed dependence of the interference induced
magnetoresistance.\cite{our1} As is seen from Fig.~\ref{fig20}, the
dependence $4\pi l^2 W(S)$ mostly follows $S^{-1}$-law for all three
systems while $S>l^2$. This is in agreement with the result of the
diffusion theory for infinite 2D system  which gives $4\pi l^2
W(S)=S^{-1}\tanh(\pi S/l^2)\simeq S^{-1}$ for
$S>l^2$.\cite{our1,samokh} Thus, despite the fact that the
distribution of scatterers is correlated in the system C, this
system shows the diffusive behavior for large number of scattering
events. A significant difference in $W(S)$ is evident for small
areas enclosed, $S\ll l^2$, where the main contribution to $W(S)$
comes from the closed paths with small number of collisions. It is
obvious that details of single scattering event and the distribution
of scatterers are important for statistics of short closed paths.
For system A the area distribution function $W(S)$ tends to the
asymptotic $S^{-1/2}$ law valid when $S\ll l^2$ [see
Refs.~\onlinecite{dyak} and \onlinecite{zdu}]. For other two
systems, $W(S)$ demonstrate more complicated behavior in our
$S$-range.

So, this qualitative consideration shows that the low-field
magnetoresistance ($b\ll 1$) which is mainly determined by the paths
with large areas enclosed ($S\gg l^2$) must be similar for all three
systems. Strong difference should appear only in high magnetic field
($b\gg 1$) when short closed paths with $S\ll l^2$ are suspended
from the interference.

Now we are in position to discuss the magnetoconductance
$\Delta\sigma(b)=\sigma(b)-\sigma(0)$. Namely this quantity is is
usually measured to obtain the phase relaxation time experimentally.
For this purpose an experimental $\Delta\sigma$-vs-$b$ plot are
fitted to the one of the known theoretical expressions with $\tau
_{\phi }$ as fitting parameter. The formula\cite{Hik} is widely used
for this purpose

\begin{equation}
\Delta\sigma(b)=\alpha G_0 \biggl\{
\psi\left(\frac{1}{2}+\frac{\gamma }{b}\right)
- \psi\left(\frac{1}{2}+\frac{%
1}{b}\right)- \ln{\gamma} \biggr\}, \label{eq6}
\end{equation}
where $\gamma=\tau/\tau_\phi$, $\psi(x)$ is a digamma function,
$\alpha$ is the prefactor. Theoretically, the prefactor is equal to
unity. Experimentalist  uses it as the fitting parameters together
with $\tau_\phi$ (or $\gamma$). Below we follow this way considering
our simulation data as the experimental ones.  Then, we  compare the
fitting value of $\gamma$ with $\tau/\tau_\phi$ used in the
simulation procedure.

\begin{figure}
\includegraphics[width=0.8\linewidth,clip=true]{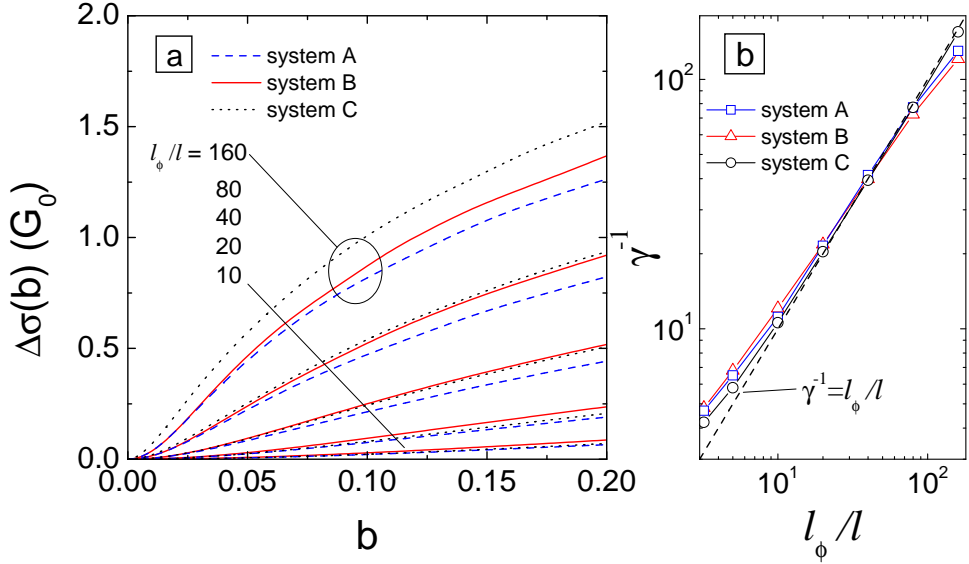}
\caption{The low-field dependence of $\Delta\sigma$ for different
$l_\phi/l$ values (a) and the results of data processing of this
curves with the help of Eq.~(\ref{eq6}) (b).} \label{fig30}
\end{figure}

Fig.~\ref{fig30}(a) shows the low-field ($b<0.2$) magnetoconductance
$\Delta\sigma(b)$ calculated  from Eq.~(\ref{eq2}) for different
$\tau_\phi/\tau$ values. As expected all  systems demonstrate close
behavior. It has been also found that taking into account
nonbackscattering processes does not practically change the magnetic
field dependence of $\Delta\sigma(b)$ for all the systems
(insensibility of low-field magnetoconductance to the
nonbackscattering processes is discussed in Refs.~\onlinecite{gorn}
and \onlinecite{our1}).

\begin{figure}
\includegraphics[width=0.9\linewidth,clip=true]{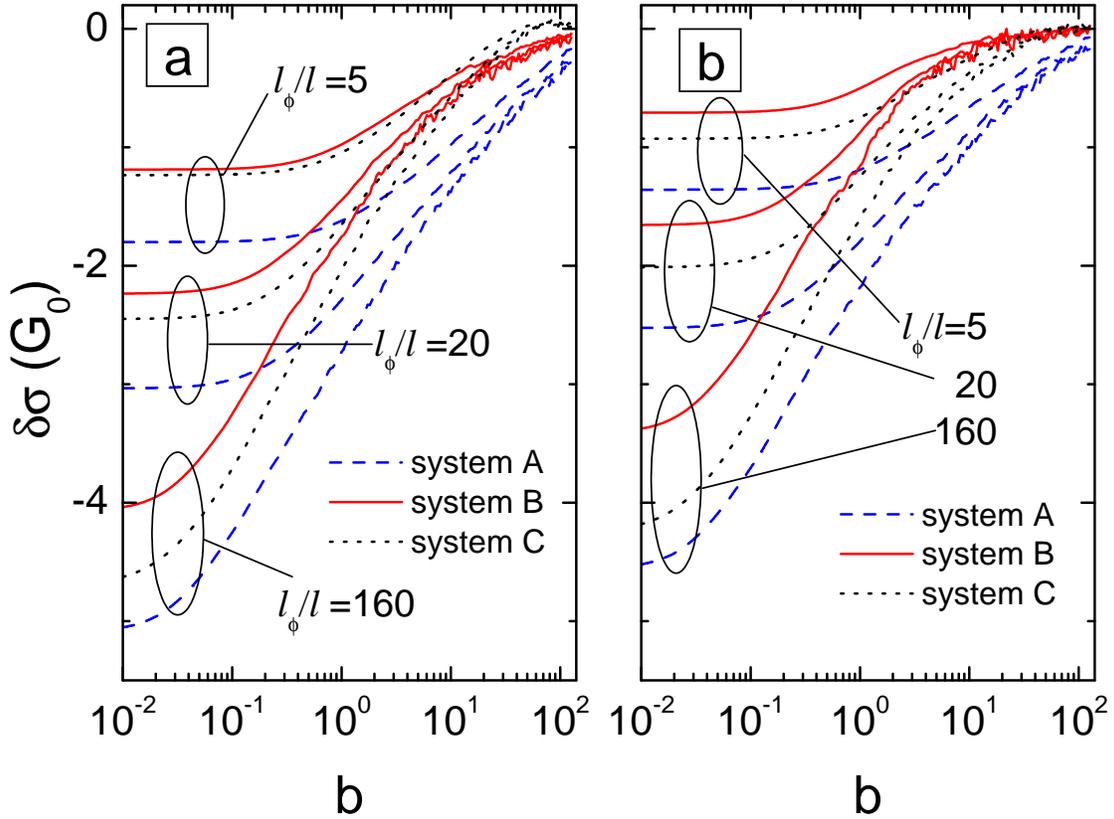}
\caption{The magnetic field dependence of $\delta\sigma$ for
different $l_\phi/l$ values, obtained with the help Eq.~(\ref{eq2})
(a) and Eq.~(\ref{eq3}) (b).} \label{fig40}
\end{figure}

If one considers the simulated $\Delta\sigma$-vs-$b$ data shown in
Fig.~\ref{fig30}(a) as experimental ones and fits them by
Eq.~(\ref{eq6}) as described above we obtain a nice coincidence in
all the cases. How the fitting value of $\gamma$ matches the value
of $l/l_\phi=\tau/\tau_\phi$ put in Eq.~(\ref{eq2}) is shown in
Fig.~\ref{fig30}(b). As seen the fitting procedure  gives the value
of $\gamma$ which can differ by a factor of about two from the value
of $l/l_\phi$. This may bee explained by the fact that
Eq.~(\ref{eq6}) is derived within the framework of the diffusion
approximation ($\gamma\ll 1$, $b\ll 1$), which is too
strong.\cite{chak,schm,our1}

\begin{figure}[t]
\includegraphics[width=0.9\linewidth,clip=true]{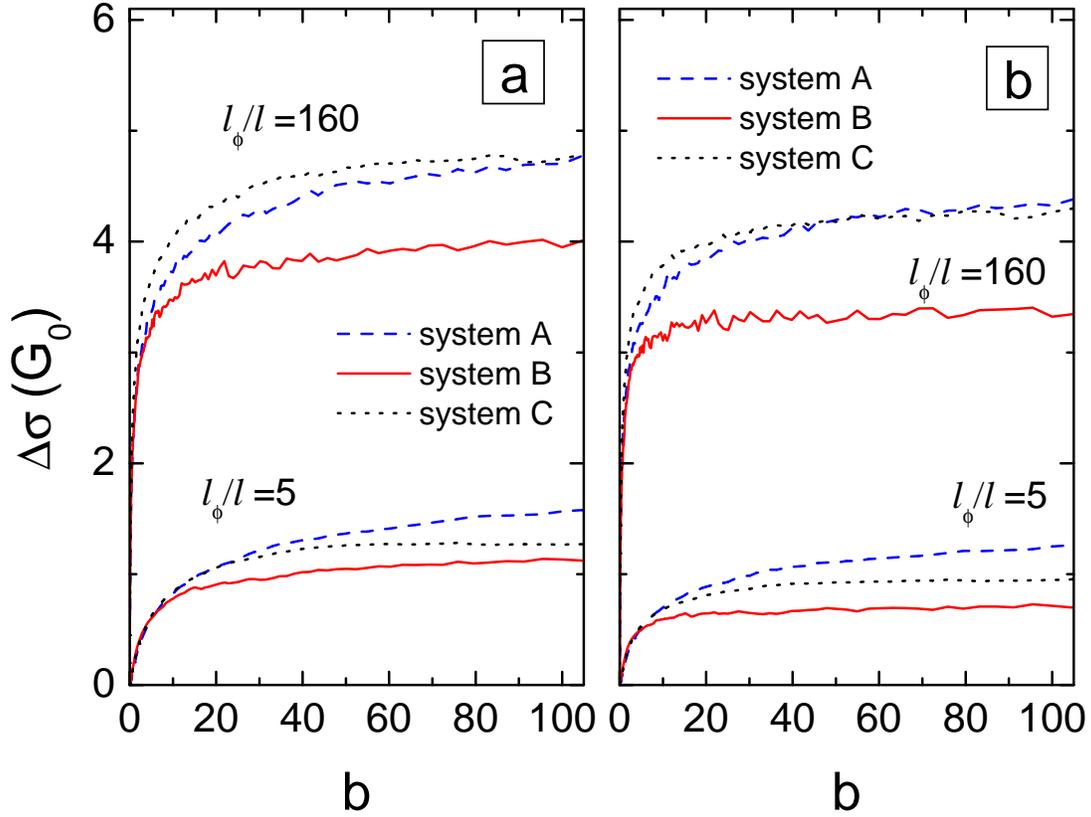}
\caption{The magnetic field dependence of
$\Delta\sigma=\delta\sigma(b)-\delta\sigma(0)$ for different
$l_\phi/l$ values, obtained using Eq.~(\ref{eq2}) (a) and
Eq.~(\ref{eq3}) (b).} \label{fig50}
\end{figure}

Figure \ref{fig40} shows the dependencies of $\delta\sigma(b)$,
calculated for different $\tau/\tau_\phi$ values with the help of
Eqs.~(\ref{eq2})  and (\ref{eq3}) in wider range of magnetic field
including the ballistic regime $b\gg 1$. It is  seen that  the
behavior of $\delta\sigma(b)$ for $b>1$ strongly depends on
scattering details for both types of calculations. It is evident
much better when considering the $\Delta\sigma$-vs-$b$ data
(Fig.~\ref{fig50}). So the magnetic field dependences of
$\delta\sigma$ and $\Delta\sigma$ are saturated in lower magnetic
field when the scattering is anisotropic or distribution of
scatterers is correlated. Such peculiarities directly follow  from
the difference in the area distribution functions at $S< l^2$ (see
Fig.~\ref{fig20}). There is a appreciable deficit of the paths with
small area enclosed, $S\lesssim 200$, in systems B and C as compared
with system A.

In conclusion, the magnetic filed behavior of the low-filed
magnetoconductance in two-dimensional systems due to suppression of
the quantum interference  is not universal beyond the diffusion
regime ($b\gg 1$ or $\tau/\tau_\phi\sim 1$). It strongly depends on
the scattering details. The use of  theoretical expressions which
were derived for the 2D systems with isotropic scattering and the
random distribution of the scatterers for interpretation of the
experimental data beyond the diffusion regime,  that is a typical
situation for experiments on high-mobility 2D heterostructures, can
give inadequate information on the value and temperature dependence
of the phase relaxation time.

We would like to thank Igor Gornyi for a number of illuminating
discussions. This work was supported in part by the RFBR (Grants
04-02-16626, 05-02-16413, and 06-02-16292), the CRDF (Grants
EK-005-X1 and Y1-P-05-11), and by the Grand of President of Russian
Federation for young scientists MK-1778.2205.2.


\begin{thebibliography}{9}
\bibitem{AA85}
B.~L.~Altshuler and A.~G.~Aronov, in {\em Electron-Electron
Interaction in Disordered Systems}, Edited by A. L. Efros and M.
Pollak (North  Holland, Amsterdam, 1985).

\bibitem{McPh}  S. McPhail, C.~ E. Yasin, A.~R. Hamilton, M.~Y.
Simmons, E.~H. Linfield, M. Pepper, and D.~A.  Ritchie,  Phys. Rev.
B {\bf 70}, 245311 (2004).

\bibitem{gork} L.~P. Gorkov, A.~I. Larkin, and D.~E. Khmelnitskii,  Pis'ma Zh.
Eksp. Teor. Fiz. {\bf 30}, 248 (1979) [JETP Lett {\bf 30}, 248
(1979)].
\bibitem{chak} S. Chakravarty and A. Schmid,  Phys. Rep. {\bf 140},
193 (1986).

\bibitem{dyak}  M.~I. Dyakonov, Solid State Commun. {\bf 92}, 711
(1994).

\bibitem{gorn} A.~P. Dmitriev, V.~Yu. Kachorovskii, and I.~V. Gornyi,
Phys. Rev. B {\bf 56}, 9910 (1997).


\bibitem{schm} H.~P. Wittman  and A.~Schmid, J. Low. Temp. Phys. {\bf 69},
131 (1987).

\bibitem{our1} G.~M. Minkov, A.~V. Germanenko, V.~A. Larionova, S.~A. Negashev, and I.~V.
Gornyi,  Phys. Rev. B {\bf 61}, 13164 (2000).


\bibitem{samokh} K.~V. Samokhin, Phys. Rev. E {\bf 59}, R2501
(1999).

\bibitem{zdu} A.~Zduniak, M.~I.~Dyakonov, and W.~Knap, Phys. Rev. B {\bf 56}, 1996 (1997).

\bibitem{Hik} S. Hikami, A. Larkin, and Y.~Nagaoka, Prog. Theor. Phys. {\bf
63}, 707 (1980).





\end{thebibliography}
\end{document}